\documentclass[]{aa}  
\usepackage{ulem}
\usepackage{graphicx}
\usepackage{txfonts}
\begin{document}

   \title{Rapid Optical Flare in the Extreme TeV Blazar 1ES 0229+200 on Intraday Timescale with TESS}

   \author{Shubham Kishore\inst{1,2}
          \and
          Alok C.\ Gupta\inst{1,3} 
          \and
          Paul J.\ Wiita\inst{4}
          \and
          S.\ N.\ Tiwari\inst{2}
    }
   \institute{Aryabhatta Research Institute of Observational Sciences (ARIES), Manora Peak, Nainital 263001, India
         \and
             Department of Physics, DDU Gorakhpur University, Gorakhpur - 273009, India
        \and 
        Xinjiang Astronomical Observatory, Chinese Academy of Sciences (CAS), 150 Science-1 Street, Urumqi 830011, PR China
        \and
        Department of Physics, The College of New Jersey, 2000 Pennington Rd., Ewing, NJ 08628-0718, USA \\
        \\
        \email{amp700151@gmail.com (SK), acgupta30@gmail.com (ACG), wiitap@tcnj.edu (PJW)}}

  \abstract
   {The extreme TeV blazar 1ES 0229+200 is a high-frequency-peaked BL Lacertae object. It has not shown intraday variability in extensive optical and X-ray observations. Nor has it shown any significant variability on any measurable timescale in the 1-–100 GeV energy range over a 14-year span, but variations in the source flux around its average are present in the energy range above 200 GeV.} 
   {We searched for intraday optical variability in 1ES 0229+200 as part of an ongoing project to search for variability and quasi-periodic oscillations in the high-cadence (2 minutes), nearly uniformly sampled optical light curves of blazars provided by the Transiting Exoplanet Survey Satellite (TESS).}
   {1ES 0229+200 was monitored by TESS in its Sectors 42, 43, and 44. We analysed the data of all these three sectors both with the TESS provided {\tt lightkurve} software and the {\tt eleanor} reduction pipeline. We detected a strong, essentially symmetric flare that lasted about 6 hours in Sector 42. We fit the flare's rising and declining phases to exponential functions. We also analysed the light curve of Sector 42  using the Lomb-Scargle periodogram (LSP) and continuous auto-regressive moving average (CARMA) methods.}
   {The optical light curve of Sector 42 of the TESS observations displayed in the present work provides the first evidence of a strong, rapid, short-lived optical flare on the intraday timescale in the TeV blazar 1ES 0229+200. 
   The variability timescale of the flare provides the upper limit for the size of the emission region to be within (3.3$\pm$0.2 -- 8.3$\pm$0.5) $\times$ 10$^{15}$ cm. Away from the flare, the slope of the periodogram's power spectrum is fairly typical of many blazars ($\alpha < 2$), but the nominal slopes for the flaring regions are very steep ($\alpha \sim$ 4.3), which may indicate the electron distribution undergoes a sudden change.
   We discuss possible emission mechanisms that could explain this substantial and rapid flare.}
   {}

   \maketitle
\nolinenumbers
\section{Introduction}
The investigation of rapid flux variability offers a strong probe into the physical mechanisms of the innermost regions of active galactic nuclei (AGNs). The blazar subclass of AGNs has stochastic flux variations in the light curves (LCs), which exhibit the largest magnitude changes as well as those over the shortest timescales. 
The observed variation in blazars LCs have been used to deduce several parameters of underlying physical processes that are not directly perceivable \citep[e.g.][and references therein]{2021ApJ...909..103Z,2023ApJS..265...14R,2023MNRAS.522..102R,2024MNRAS.527.1344D}.
However, it can be quite difficult to explain the physical and astrophysical processes associated with the normally aperiodic nature of the LCs of blazar since they have been seen to vary on all timescales, ranging from a few minutes to several years, in all observable electromagnetic (EM) bands.  Blazar variability from GeV observations with Fermi-LAT can often be characterised by the exponential Ornstein-Uhlenbeck process \citep{2021A&A...645A..62B}. Perhaps the most puzzling blazar variability is that which occurs on timescales ranging from a few minutes to less than a day, often called intraday variability (IDV) \citep{1995ARA&A..33..163W}. Multiple methods, including determination of variability timescales, power spectrum density (PSD) analysis, regression methods, correlation methods, etc., have been used to gain insight into these objects \citep[e.g.][and references therein]{2020ApJS..250....1T,2023MNRAS.522..102R,2023ApJ...957L..11G}. In this context, it generally appears that for blazars, the sub-class of high energy/synchrotron peaking blazars (HBLs/HSPs) show lower amplitude variability with a lower duty cycle (DC) in the optical bands \citep[e.g.][and references therein]{2012MNRAS.420.3147G,2012AJ....143...23G,2016MNRAS.458.1127G,2019ApJ...871..192P,2020MNRAS.496.1430P,2020ApJ...890...72P,2023MNRAS.519.2796D,2024MNRAS.527.1344D}. The opposite seems to be usually true for the low energy/synchrotron peaking blazars (LBLs/LSPs) \citep[e.g.][and references therein]{1989Natur.337..627M,1996A&A...305...42H,1999A&AS..134..453S,2008AJ....135.1384G,2009ApJS..185..511P,2011MNRAS.413.2157R,2012MNRAS.424.2625B,2023MNRAS.522.3018B,2012MNRAS.425.3002G,2024MNRAS.527.5220T}. \\ 
\\
The blazar 1ES 0229+200 was first detected in the Einstein Imaging Proportional Counter (IPC) Slew Survey \citep{1992ApJS...80..257E} and subsequently classified as a BL Lacertae object \citep{1993ApJ...412..541S}. It has a red-shift of $z = 0.1396$, and the mass of its central supermassive black hole (SMBH) has been estimated to be $\sim$4.8 $\times$ 10$^{8}$ M$_{\odot}$ \citep{2005ApJ...631..762W}. Based on its high X-ray-to-radio flux ratio, it was classified as an HBL \citep{1995A&AS..109..267G}. 1ES 0229+200 is among a few blazars which were extensively observed in the first 5.5 months of observations by the Fermi-LAT \citep{2009ApJ...707.1310A}. It is listed in the catalogue of TeV emitting sources\footnote{\url{http://tevcat.uchicago.edu/}}.  The VLA radio observations of this blazar demonstrate curved jets and found a core flux of 51.8 mJy at 5 GHz \citep{2003AJ....125.1060R}. In 184 nights of optical observations over $\sim$5 years (2007--2012), its brightness varied only slightly, within $\approx$0.2 magnitudes, and no clear colour magnitude correlation was found \citep{2015A&A...573A..69W}. In a recent X-ray polarisation study of the source in the energy range 2--8 keV with the Imaging X-ray Polarimetry Explorer (IXPE), a degree of polarisation of 17.9\% $\pm$ 2.8\% and an electric-vector position angle of 25$^{\circ}$.0 $\pm$ 4$^{\circ}$.6 were found \citep{2023ApJ...959...61E}. \\

\begin{figure*}
    \resizebox{18.2cm}{!}{\includegraphics[trim=1.cm 0 0 .5cm]{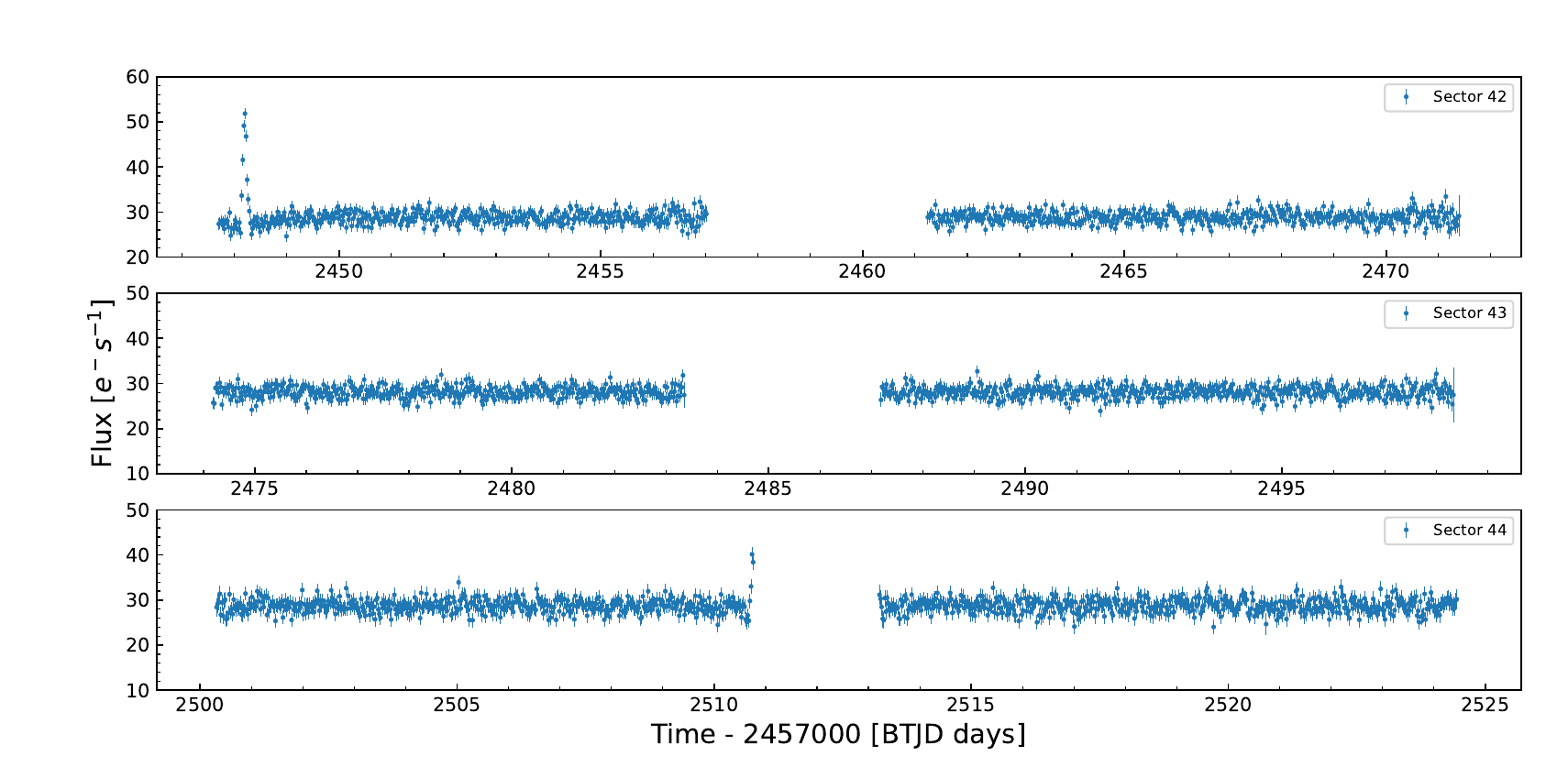}}
    \vspace{-.3cm}
    \caption{Sectoral TESS LCs of 1ES 0229+200 after data reduction  (in 30-minute bins for better visualisation.)}
 
    \label{sect_LC}
\end{figure*}

\noindent
1ES 0229+200 has shown some very peculiar behavior. Very high energy (VHE) TeV $\gamma-$ray emission from the source was detected, with a spectrum characterised by a hard power-law in the energy range of 0.5--15 TeV \citep{2007A&A...475L...9A}.  While those early data showed no evidence for significant TeV variability, additional VHE observations have shown variations in flux on timescales of months \citep[see][]{2023A&A...670A.145A}. It is considered as an extreme TeV emitting blazar in both its synchrotron and Compton emissions in the sense that its  X-ray emission is detected up to $\sim$100 keV with a very hard spectrum (with photon index $\Gamma \sim 1.8$) showing an excess absorption above the Galactic value \citep{2011A&A...534A.130K}. \cite{2011A&A...534A.130K} also mention it as having one of the highest inverse Compton (IC) peak frequencies known at that time. The multi-wavelength (MW) spectral energy distribution (SED) of 1ES 0229+200 is well fitted by an SSC (synchrotron self Compton) model with a very high Doppler factor in the range of 40--100 \citep{2009MNRAS.399L..59T,2011A&A...534A.130K,2014ApJ...782...13A}. In 14 nights of optical R band monitoring during 2016--2019, no IDV was detected in any of the nights \citep{2020MNRAS.496.1430P}. In several X-ray LCs, three made with NuStar and two with XMM-Newton, all having substantial good time intervals (GTIs) in the range of $\sim$16--21 ks, no IDV was detected \citep{2017ApJ...841..123P,2022ApJ...939...80D}. \\
\\
To better understand blazars’ optical flux variabilities on IDV time scales, we have begun using data from the Transiting Exoplanet Survey Satellite (TESS)\footnote{\url{http://tess.gsfc.nasa.gov}} \citep{2014SPIE.9143E..20R} to search for strong flaring, stochastic variability, and quasi-periodic oscillations (QPOs) in the essentially evenly sampled LCs with high cadence (down to 2 min) that TESS allows. Recently, we reported the detection of several QPOs with periods in the range of 0.6--2.5 days in the optical emission of the blazar S4 0954+658 \citep{2023ApJ...943...53K}. We have also reported on two exceptional optical flares in the blazar OJ 287, with fluxes nearly doubling and then nearly tripling over two days \citep{2024ApJ...960...11K}. In the present work, we report for the first time the detection of a strong optical flare on an IDV timescale in the blazar 1ES 0229+200. As described above, until now, observations of this source have not shown much variability on IDV timescales in any EM band. \\
\\
In Section 2 of this Letter, we briefly describe our TESS data analysis. In Section 3, we describe the LC analysis and results. A discussion and our conclusions are given in Section 4.

\section{Observations and Data Reduction}
TESS observed the blazar 1ES 0229+200 in three consecutive sectors (42, 43, and 44) spanning 2021 August 21 to 2021 November 05 at a cadence of 2 minutes. Each sector typically lasts 27 days, and there are  1--3 day-long inter- and intra- sectorial gaps in the data during which the satellite sends the observed data to Earth or waits for any commands to be uploaded. \\

\begin{figure*}
       \centering
       \resizebox{19.cm}{!}
       {\includegraphics[trim=0 1cm 0 0]{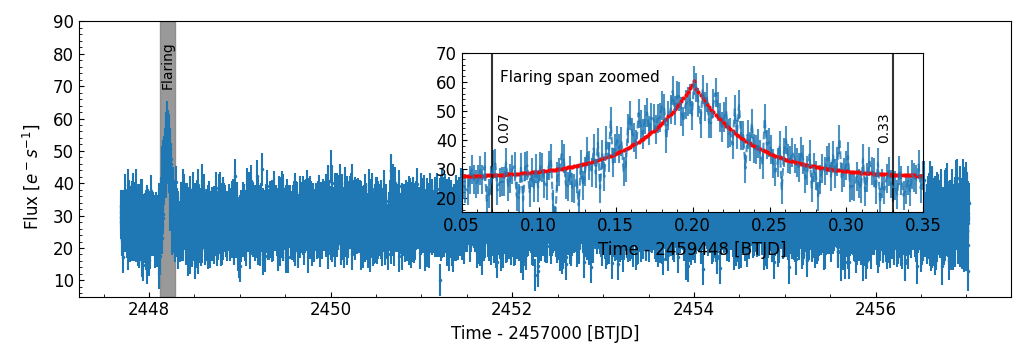}}
    \resizebox{19cm}{!}{\includegraphics{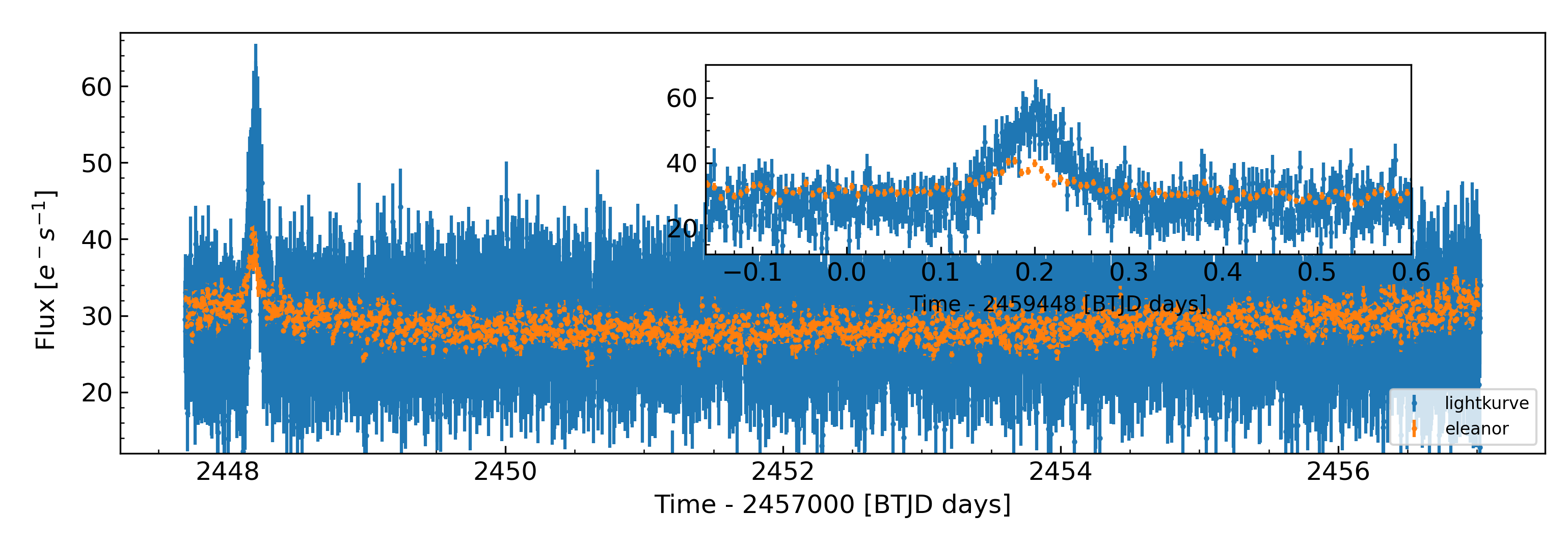}}
    
    \caption{The upper plot includes the complete reduced Sector 42 LC using {\tt lightkurve} where the inset zooms in on the flare period. The red curve in the subplot is the model fitted by Eq.\ \ref{e_folding}. The lower plot presents a comparison between the LCs produced using {\tt lightkurve} and {\tt eleanor}. The {\tt eleanor} LC in this plot has been shifted down by 40 e$^-$s$^{-1}$. A zoomed view of the flare with the {\tt eleanor} reduction is plotted in the inset.}
    \label{flaring_fit}
\end{figure*}

\begin{figure*}
    \resizebox{18.5cm}{!}{\includegraphics{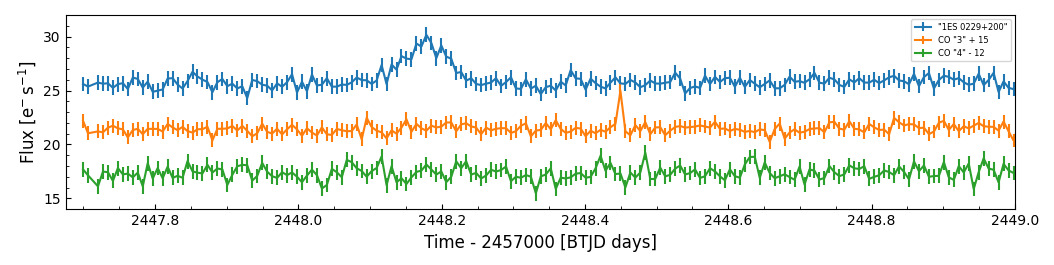}}
    \vspace{-.8cm}
    \caption{Partial Sector 42 median flux LCs of 1ES 0229+200 blue and comparison objects (CO) `3' orange) and `4' (green). It should be emphasised that this plot provides a preliminary version of the LC (obtained from the {\tt targetpixelfile created using {\tt TESSCUT}} for each of the individual objects indicated).  While this provides a useful visual comparison, the more fully reduced version of {\tt PDCSAP\_flux} from the already created {\tt lightcurvefile} has been used in our actual analysis. The comparison had to be conducted this way as there were no {\tt lightcurvefile} outputs available for the COs.}
    \label{comp_LC}
\end{figure*}

\noindent
We have used and reduced the PDCSAP\_FLUX \citep{2016SPIE.9913E..3EJ} data for our analysis, which is provided by TESS with a cadence of 2 minutes, in the same manner as in \citet{2023ApJ...943...53K,2024ApJ...960...11K}. We refer the readers to these papers for details of this approach using Cotrending Basis Vectors (CBV) and a discussion of the reduction procedure using the TESS {\tt lightkurve} software.  The goal here is to have the highest possible values of both overfitting and underfitting metrics consistent with a low value of the regularisation parameter $\alpha$.   Table \ref{tab:cal.details} includes the values of the fitting parameters obtained during the reduction of PDCSAP\_FLUX of our object for each of the sectors, indicating that this goal was achieved for each of these observations of 1ES 0229+200.  Fig.\ \ref{sect_LC} shows these data in all three sectors reduced in this fashion.\\
\\
Other groups have designed reduction pipelines for TESS data that aim to produce LCs for AGN that take better account of the variable backgrounds and instrumental noise.  These pipelines include {\tt eleanor} \citep{2019PASP..131i4502F,2019ascl.soft05007B} and {\tt quaver} \citep{2023ApJ...958..188S}. Recently \citep{2024arXiv240314744P} have conducted a careful comparison of three methods based on the {\tt lightkurve} package (CBV, Regression, and Pixel Level Decorrelation) as well as the  {\tt eleanor}  \citep{2019PASP..131i4502F} and {\tt quaver} \citep{2023ApJ...958..188S} pipelines and the simple differential photometry reduction method.  The key result of this comparison is that the three latter methods provide better matches to simultaneous ground-based observations of several blazars. While the direct methods using {\tt lightkurve} usually show the same types of variations in the LCs as do the pipelines, they tend to squash the amplitude of the variability. \\
\\
Therefore, we have also used the {\tt eleanor}  python package to check the results produced with {\tt lightkurve} where the reduced {\tt pca\_flux} was used for analysis. The baseline flux for the {\tt eleanor} LC was found to be shifted higher by $\sim$$\rm{40} \ e^{-} s^{-1}$, and we note that {\tt eleanor} uses the longer cadence of 10\ minutes.  A comparison of the LCs produced with these two different methods is shown in Fig.\ \ref{flaring_fit}. 
The variation of the baseline could be due to differences in background estimation, hence leading to the visual long-term variation. Although we do not have simultaneous ground-based observations that could decide which approach is better here, we note that the flare is clearly seen in both LCs.  And {\it contra} Poore et al.\ (2024), in this case, the relative amplitude of the flare is actually smaller for {\tt eleanor}.  
 
\begin{table}[h]
\caption{Flux calibration details}
\begin{tabular}{c c c c}\hline\hline
 Sector & $\alpha$ & Overfitting metric & Underfitting metric\\\hline
 42 & 0.10 & 0.996 & 1.000 \\
 43 & 0.10 & 0.998 & 1.000 \\
 44 & 0.10 & 0.996 & 1.000 \\
\hline
\end{tabular}
\label{tab:cal.details}
\end{table}

\section{Light Curve Analysis and Results}
Visual inspection of the LC of Sector 42 of the blazar 1ES 0229+200 in Fig.\ \ref{sect_LC} gives clear evidence of strong but short-lived, rapid variability. On the other hand, the LCs of two other epochs observed in Sectors 43 and 44 do not show any obvious variation. To confirm the genuine nature of the variability in the LC of Sector 42, we used the TESS LCs of comparison stars `3' and `4' from the USNO 2.0 Catalogue \citep{1998AAS...19312003M}\footnote{\href{htps://www.lsw.uni-heidelberg.de/projects/extragalactic/charts/0229+200.html} {Field-of-view for 1ES 0229+200}} to investigate the LC behaviors near the epoch of apparent high variability of the source in the same manner as in \citet{2024ApJ...960...11K}. To perform this comparison, we took a $10\times10$ cutout of the full-frame images of the source blazar and the noted comparison stars. Using the {\tt aperture} module available in the {\tt lightkurve}  package, we extracted the median flux LC of all these objects. A background of $\sim$$\rm{151} \ e^{-} s^{-1}$ in the cutouts was found around the time of the flare. We plot the background subtracted median flux LCs of these objects in Fig.\ \ref{comp_LC}. The increased flux of the blazar over many consecutive data points, along with simultaneous relatively quiescent behavior of the comparison objects (COs) (with all nominal variations spanning very few points), clearly indicates an intrinsic variability of the blazar. A zoomed portion of this high variability region is shown in Fig.\ \ref{flaring_fit}, revealing a rapid flare centered at BTJD 2448.20. It should be noted that these median flux values have only been used here to make clear the presence of the flare.\\
\\
Our detailed analysis requires the determination of the baseline flux both before and after the flare, the starting and ending epochs and the e-folding timescales.  The latter were separately fitted for both the rising and declining phases of the flare,  using the least square method, with the model found to be
\begin{equation}
F(t)=exp \ \Big(\frac{t-t_0}{\tau}\Big)+c~.
\label{e_folding}
\end{equation}
Here $c$ is the baseline flux, and $\tau$ gives the e-folding timescale. In the rising phase of the flare, $t_0$ gives the time of onset of the flare, while in the declining phase, it is the flare ending time; with this set of parameters, no additional normalisation is required for the exponential terms. Table \ref{Flare_characteristics} includes the parameters individually found for the two phases of the flare using our direct {\tt lightkurve} CBV approach, and the upper panel of Fig.\ \ref{flaring_fit} explicitly shows the flaring spans along with the model fit. These indicate that the flare is essentially symmetric.  The flare reaches a maximum flux of $\sim$60 e$^-$s$^{-1}$ at the epoch $\sim$2448.201\ d with an associated uncertainty (which happens to be maximum during nearby epochs) of $\sim$5\ e$^-$s$^{-1}$. Taking a mean baseline of 27\ e$^-$s$^{-1}$ from the obtained values and using the maximum local uncertainty, this flare exhibits a $\sim $6.6$\sigma$ ((max. flux\ --\ baseline)/uncertainty) significance level detection, making it quite an interesting phenomenon for 1ES 0229+200, which has otherwise been observed to be quiet on short timescales. \\
\\
We also show the LC from {\tt eleanor} in the lower panel of Fig.\ \ref{flaring_fit} and note the excellent agreement of the timescales obtained for the rising part of the flare, which is $0.8\pm0.1$\ h here. However, using {\tt eleanor}, the LC showed a more relaxed decay timescale of $\sim$1.9\ h. The confirmation of the rising timescale is helpful in determining the size of the emission region, assuming this value for the most probable shortest variability timescale \citep[see, e.g.,][]{2024ApJ...960...11K}.
  
\begin{table}
\caption[]{Flare characteristics}
\label{Flare_characteristics}

\resizebox{3.5in}{!}{
\begin{tabular}{c c c c}     
\hline\hline                 
    &$c$ ($e^-s^{-1}$)& $t_0 - 2457000$ (BTJD)& $\tau$ (d)\\
    \hline                 
Rising phase & 26.9$ \pm$ 0.3 & 2448.07 $\pm$ 0.01 & 0.037 $\pm$ 0.003 \\
Declining phase& 27.2 $\pm$ 0.3 & 2448.33 $\pm$ 0.01 & $-$0.036 $\pm$ 0.003 \\
\hline                       

\hline                                  
\end{tabular}}
   \end{table}

\begin{figure*}[t]
    
    \resizebox{19cm}{!}{
    \includegraphics[trim=0 1cm 0 0]{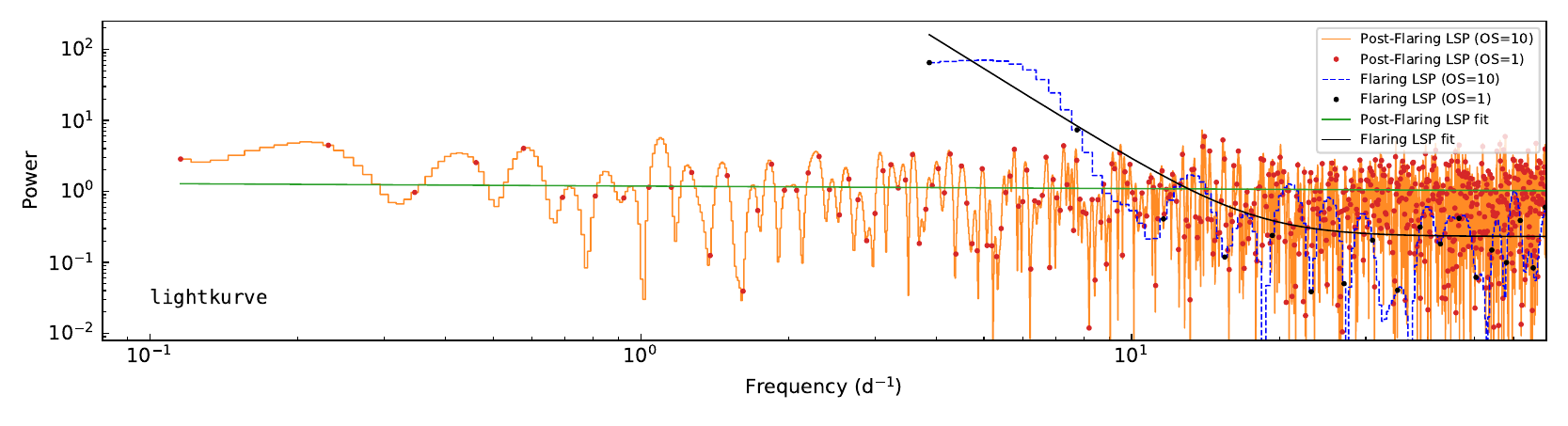}}
    \resizebox{19cm}{!}{
  \includegraphics{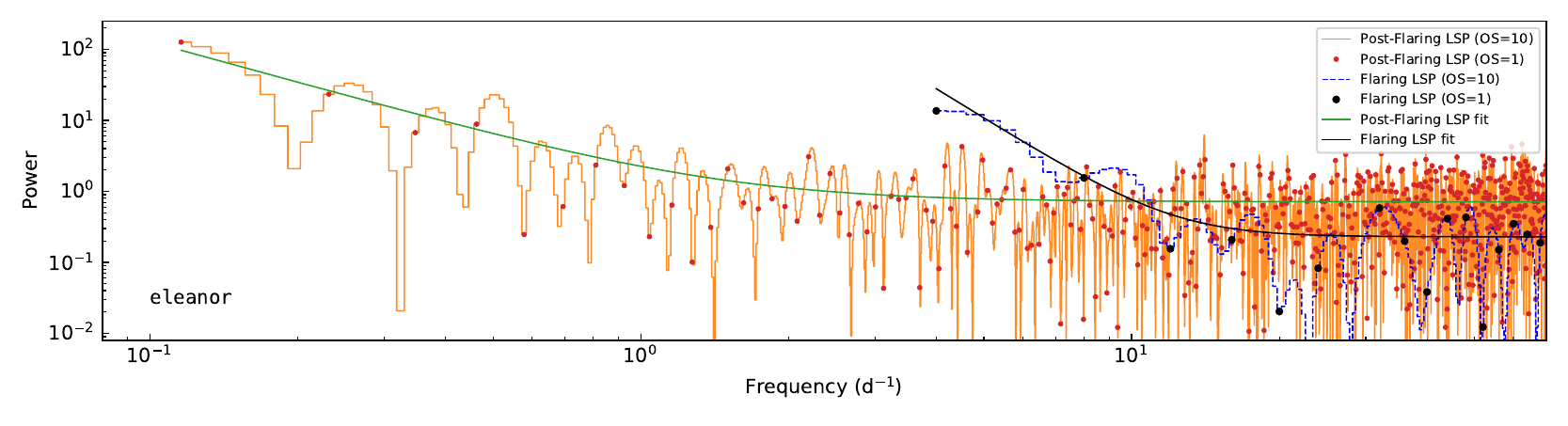}}
    \caption{LSP plots of flaring and post-flaring portions of the first segment of the Sector 42 LC. The upper plot shows LSP corresponding to the LC achieved with {\tt lightkurve} while the lower plot corresponds to the {\tt eleanor}. In these plots, the LSPs are plotted corresponding to the oversamplings (OS) indicated. The normal LSPs with no oversampling are over-plotted for demonstrating the unreliability of their fitting due to the scarcity of available data points. The LSP fits correspond to OS=10}
    \label{fig:LSP_plots}
\end{figure*}

\subsection{Temporal Analysis}
The periodogram of a LC gives an estimate of the strengths of various temporal frequencies in the LC. Typical blazar periodograms are characterised by a red noise power law spectrum at lower frequencies, which at high frequencies becomes asymptotic to a white noise \citep[e.g. see][and references therein]{2014ApJ...785...60R,2019ApJ...877..151W,2024ApJ...960...11K}. A highly dominant frequency compared to the background spectrum implies a periodic/quasi-periodic oscillation; however, even in the absence of such peaks in the periodogram, its spectral index may yield information about the source of variability since different physical models naturally produce distinct ranges of spectral index.\\
\\
The parameters shown in Table \ref{Flare_characteristics}, obtained from fitting Eq.\ \ref{e_folding}, were used to separate the flaring region from the rest of the first segment of Sector 42 LC, as shown in Fig.\ \ref{flaring_fit}. Both the post-flare and flare portions of that LC were then individually analysed to estimate the spectral index in the frequency domain using a Lomb-Scargle periodogram (LSP) with `HorneBaliunas' normalisation \citep{1976Ap&SS..39..447L, 1982ApJ...263..835S, 1986ApJ...302..757H}. As always, there is a sparsity of data points in the low-frequency range of the LSP, and this can substantially affect the fitting of the periodogram, so we employed an oversampling factor of 10 while calculating the LSPs. The two LSPs were fitted using a log-likelihood method \citep[see Eq.\ 17 of][]{2010MNRAS.402..307V}  with a simple power-law model as well as the associated white noise: 
\begin{equation}
    P(\nu)=A\nu^{-\alpha}+C~.
    \label{Power law}
\end{equation}
The upper panel of Fig.\ \ref{fig:LSP_plots} shows the two LSPs obtained and their corresponding model fits for our standard approach, while the LSP fitting parameters are listed in the first two rows of Table\ \ref{LSP_comp_tab}. The parameter of interest here is $\alpha$, which is    $\sim$0.04 (only simple power law preferred)
for the post-flare bulk of the segment's LC but is nominally very steep ($\sim$4.3) for the flare portion.  We do note, however, that the flare PSD is being fit over quite a limited range in frequency thanks to the modest duration of the flare, so this steep $\alpha$ is a tentative result.  In the lower panel of Fig.\ \ref{fig:LSP_plots}, we show the PSDs found using the {\tt eleanor} LC. The long-term variation visible (probably due to a different background subtraction) in the lower plot of Fig.\ \ref{flaring_fit} may well account for the greater power in the lower frequency range of this LSP (lower plot of Fig.\ \ref{fig:LSP_plots}) which leads to a steeper slope of $\sim$1.9.  The nominal value for the flare's PSD  slope was $\sim$4.3, which is in agreement with the steep slope obtained with {\tt lightkurve}.   
To investigate the impact of any uncertainty in the appropriate length of flare on the PSD plots we considered several different temporal extents for the flare and corresponding different lengths for the post-flare dataset, using our {\tt lightkurve} reduction.  In all cases the flare PSD slopes exceeded 4 while the post-flare slopes were less than 1.
 Though our primary focus is on the first segment of the Sector\ 42 light curve, we analysed other segments of other sectors too made using the {\tt lightkurve} reduction, and found that their LSPs are all consistent with only white noise. 

\begin{table}
\caption[]{Comparison of flaring and post-flaring  LSP parameters}
\label{LSP_comp_tab}
\centering
\resizebox{3.5in}{!}{
\begin{tabular}{c c c c c}  
\hline\hline
Reduction method& LSP & $log_{10}(A)$ & $\alpha$& C \\
\hline      
lightkurve&Flaring  & 4.72 & 4.29 & 0.23 \\
&Post-flaring (sec. 42/1) & 0.07 & 0.04 &  --\\
eleanor&Flaring & 4.03 & 4.30 & 0.23 \\
&Post-flaring  (sec. 42/1)  & 0.19 &  1.92 & 0.71\\
\hline   
\end{tabular}}
   \end{table}

\subsection{CARMA Analysis}
\noindent
A somewhat more sophisticated method to analyse the LC makes the initial assumption that it is a manifestation of a Gaussian noise process. The continuous autoregressive moving average (CARMA) method, based on this assumption, describes the LC and its first $p$ derivatives in terms of the underlying noise and its first $q$ derivatives with respect to time; and is hence assisted with two parameters, p and q, denoted as CARMA$(p,q)$ \citep[e.g.][]{2014ApJ...788...33K}. The CARMA models essentially describe the connection between the short-term memory of the process and the behavior of random fluctuations on different timescales. \\
\\
We also analysed the flaring and non-flaring portions of the Sector 42 LC reduced with the CBV approach using CARMA, following the method of \cite{Yu2022}. We considered all the $(p,q)$ pairs with $1 \leq p \leq 5$ and $q < p$ for the CARMA fittings and the best $(p,q)$ set was estimated by minimizing the negative of the log-likelihood found for all those $(p,q)$ sets. We found that both portions were preferably fitted with the simplest CARMA(1,0) model; however, the likelihood of a CARMA(2,0) model was very similar to that of the CARMA (1,0) model for the non-flaring portion. 

\section{Discussion and Conclusions}
\noindent
In this paper, we have considered the LC of three sectors (42--44) of the TESS observations of the blazar 1ES 0229+200. The object shows almost no optical variations during that period except for a flare in the first segment of Sector 42. We find that the flare is highly symmetric, exhibiting identical e-folding timescales within errors. This is evidence that this variability is regulated by the radiation/disturbance crossing time through the emission region instead of the acceleration or energy-loss timescales of the radiating electrons \citep[see][]{1999MNRAS.306..551C,2019MNRAS.482..743R}. Multiple evidences of sharp hour-time-scale flares, with rises and decays that are approximately linear or exponential, have been found over the past years for different blazars in different EM-bands \citep[e.g.][and references therein]{1993A&A...271..344W,2008Natur.452..966M,2012ApJ...749..191C,2013ApJ...766L..11S,2014ApJ...796...61K,2024ApJ...960...11K}.\\
\\
The simple causality constraint given as
\begin{equation}
    R \leq \frac{c ~\tau ~\delta}{1+z} ~,
\end{equation}
can be utilised to limit the size of emission region, $R$, where $c$, $\tau$, $\delta$ and $z$ are the speed of light, variability timescale, Doppler factor and red-shift, respectively. Very high Doppler factors  with a large range of 40--100 for 1ES 0229+200 have been estimated via MW SED modeling \citep{2009MNRAS.399L..59T,2011A&A...534A.130K,2014ApJ...782...13A}. Taking $z=0.1396$ and considering an average e-folding timescale  ($\sim$0.036$\pm$0.002\ d, or $\sim$0.88$\pm$0.05\ h) as the variability timescale, for this range of $\delta$, we find the upper limit to the size of the emission region to be within $(3.3\pm0.2 - 8.3\pm0.5)\times10^{15}$\ cm, assuming that the uncertainty is propagated via variability timescale only. \\
\\
Properties of the variability, such as the timescale and periodogram, can provide significant insight into the driving process. Different proposed models spontaneously result to specific variability timescales and the spectral indices of their PSDs.  
A strong, rapid, and rare flare, such as seen here for 1ES 0229+200, indicates that some extreme change has occurred in the emission region. The periodogram analysis of the flare in Sector 42 suggests a very steep PSD index, $\alpha\sim4.3$,  regardless of reduction method, compared to values typically seen ($1.4 - 3.0$) for optical variability from most blazars \citep[e.g. see][and references therein]{2012ApJ...749..191C,2018A&A...620A.185N,2019ApJ...877..151W,2020ApJ...903..134C,2021ApJ...909...39G,2023ApJ...951...58W}.  The weak variations in the  post-flare portion of that Sector have  shallower PSD slopes ($\alpha\sim$ 0.04 or 1.9, depending on the reduction method), indicating a sudden change in electron distribution during the flare.  Previously, \cite{2019ApJ...877..151W} found nominal PSD slopes of 3 (out of 9 studied) gamma-ray detected blazars in the range 3 -- 4 for K2 extended Kepler emission LCs, but they were measured on $\sim$2 -- 3 month timescales. A subsequent reanalysis of those and additional data from 40 K2 LCs measured for 29 AGN (including 16 BL Lacs) found only 1 slope steeper than $3$,  while the vast majority were between 1.6 and 2.6  \citep{2023ApJ...951...58W}.  A study of the power spectra of the intranight LCs of 14 bright blazars found a wide range of best fitting PSD slopes and a rather steep average of $2.9\pm0.3$ on those short time scales \citep{2021ApJ...909...39G}.   Over multi-year time frames, \cite{2018A&A...620A.185N} found generally shallower PSD slopes for their sample of 31 gamma-ray emitting blazar LCs, with a range from 1.0 to 1.9 and an average of $1.46\pm0.18$. So, we are unaware of any previous clear case for such steep PSD slopes of optical LC for blazars on timescales of a few hours. We note that we performed separate PSD analyses of the flare and post-flare LCs of 1ES 0229+200  while these other studies analysed the entire LCs. \\ 
\\
The EM emission in blazars is generated in relativistic jets that are beamed and Doppler-boosted towards the observer's line of sight, which dominates the thermal emission from the accretion disc, so any blazar variability observed on any timescale is more likely to be explained using relativistic jet based models. However, as there are situations when flat-spectrum radio quasars in their low states show features of disc thermal emission in their SEDs, the variability can arise from hotspots on, or instabilities in, the accretion discs \cite[e.g.][]{1993ApJ...406..420M,1993ApJ...411..602C}. Since 1ES 0229+200 is an HBL, the possibility of significant variability from the disc region can be discarded. Several jet-based models have been proposed for blazar flux variability observed on different timescales. These include: shock-in-jet scenarios \citep[e.g.][]{1985ApJ...298..114M}; turbulence behind the shocks in a relativistic jet \cite[e.g.][]{2014ApJ...780...87M,2016ApJ...820...12P};  ultra-relativistic mini-jets \cite[e.g.][]{2009MNRAS.393L..16G,2009MNRAS.395L..29G}; or turbulence produced by magnetic reconnection \citep[e.g.][]{2021ApJ...919..111G,2021ApJ...912..109K}. \\
\\
Since 1ES 0229+200 is a BL Lac, a disc emission contribution is unlikely, so the LC in Fig.\ \ref{flaring_fit} can be interpreted as a superimposition of a short-term flare in a compact magnetic reconnection region that arises from instabilities over a barely varying envelope of non-thermal jet emission.  In a similar study of $\gamma$-ray flares, \cite{2020NatCo..11.4176S} found a peak-in-peak behavior in two of the three detected fast flares of the FSRQ 3C 279 where the fast ﬂares were superimposed on the more slowly varying envelope emission; they attributed this rapid variation to a mini-jets scenario. Although their study included multiple flares, this scenario can still produce an isolated flare. Hence, we suggest this jet-in-jet or mini-jet model as a likely way to produce a rapid change in electron distribution and increased local Doppler factor that could engender the outburst seen in 1ES 0229+200 during Sector 42 of these TESS observations. 
\begin{acknowledgements}
This paper includes data collected with the TESS mission, obtained from the MAST data archive at the Space Telescope Science Institute (STScI). Funding for the TESS mission is provided by the NASA Explorer Program. STScI is operated by the Association of Universities for Research in Astronomy, Inc., under NASA contract NAS 526555. ACG's work is partially supported by the CAS ``Light of West" China Program (No. 2021-XBQNXZ-005) and the Xinjiang Tianshan Talents Program.
\end{acknowledgements} 

\noindent
{\it Facility:}  Transiting Exoplanet Survey Satellite (TESS) -- The dataset used in this paper can be found in MAST: \href{https://mast.stsci.edu/portal/Mashup/Clients/Mast/Portal.html?searchQuery=%7B%22service%22:%22DOIOBS%22,%22inputText%22:%2210.17909/b3et-af14%22%7D}{10.17909/b3et-af14.} \\
\\
{\it Software:} lightkurve \citep{2018ascl.soft12013L}, 
SciPy \citep{2020SciPy-NMeth}, eleanor \citep{2019PASP..131i4502F}, EzTao \citep{Yu2022}.

\bibliographystyle{aa}
 \bibliography{ref}

\end{document}